\documentclass[applsci,article,accept,moreauthors,pdftex,10pt,a4paper]{Definitions/mdpi} 
\pdfoutput=1
\usepackage[utf8]{inputenc}
\usepackage{graphicx}
\usepackage{amssymb,amsmath}
\usepackage{subfigure}
\usepackage{color}

\usepackage[normalem]{ulem}
\usepackage{booktabs} 
\usepackage{multirow}
\usepackage{soul} 
\usepackage{microtype}

\updates{yes}
\setenumerate{parsep=6pt,itemsep=0pt,leftmargin=*,labelsep=5.5mm}
\usepackage{upgreek} 
\setitemize{parsep=6pt,itemsep=0pt,leftmargin=*,labelsep=5.5mm}
\firstpage{1} 
\pubvolume{8}
\issuenum{0}
\articlenumber{0}
\pubyear{2018}
\copyrightyear{2018}
\history{Received: date; Accepted: date; Published: date}

\Title {A Versatile Velocity Map Ion-Electron Covariance Imaging Spectrometer for High-Intensity XUV~Experiments}

\Author{Linnea Rading %Please carefully check the accuracy of names and affiliations.Changes will not be possible after proofreading.
$^{\dagger}$, Jan Lahl *$^{,\dagger}$\orcidA{}, Sylvain Maclot $^{\dagger}$\orcidB{}, Filippo Campi, Hélène Coudert-Alteirac\orcidC{}, Bart Oostenrijk, Jasper Peschel\orcidE{}, Hampus Wikmark, Piotr Rudawski\orcidD{}, Mathieu Gisselbrecht and Per Johnsson~*\orcidF}

\AuthorNames{Linnea Rading, Jan Lahl, Sylvain Maclot, Filippo Campi, Hélène Coudert-Alteirac, Bart Oostenrijk, Jasper Peschel, Piotr Rudawski, Hampus Wikmark, Mathieu Gisselbrecht, and~Per Johnsson}

\address[1]{Department of Physics, Lund University, P.O. Box 118, 22100 Lund, Sweden;\linebreak linnea.k.rading@gmail.com (L.R.); sylvain.maclot@fysik.lth.se (S.M.); filiuqwerty@gmail.com (F.C.); helene.coudert@fysik.lth.se (H.C.-A.); bart.oostenrijk@sljus.lu.se (B.O.); jasper.peschel@fysik.lth.se (J.P.); hampus.wikmark@fysik.lth.se (H.W.); piotr.rudawski@fysik.lth.se (P.R.); mathieu.gisselbrecht@sljus.lu.se (M.G.)}

\corres{\hangafter=1 \hangindent=1.05em \hspace{-0.82em}Correspondence: jan.lahl@fysik.lth.se (J.L.); per.johnsson@fysik.lth.se (P.J.)}

\firstnote{\hangafter=1 \hangindent=1.05em \hspace{-0.82em}These authors contributed equally to this work.}

\abstract{We report on the design and performance of a velocity map imaging (VMI) spectrometer optimized for experiments using high-intensity extreme ultraviolet (XUV) sources such as laser-driven high-order harmonic generation (HHG) sources and free-electron lasers (FELs).~Typically exhibiting low repetition rates and high single-shot count rates, such~experiments do not easily lend themselves to coincident detection of photo-electrons and -ions. % suspended hyphen: "photoelectrons and -ions" = "photoelectrons and photoions"
In order to obtain molecular frame or reaction channel-specific information, one has to rely on other correlation techniques, such~as covariant detection schemes.~Our device allows for combining different photo-electron and -ion detection modes for covariance analysis.~We present the expected performance in the different detection modes and~present the first results using an intense high-order harmonic generation (HHG) source.}


\begin{document}
% swap these command definitions below to obtain the clean version without any colored marks
%\newcommand{\delete}[1]{\sout{#1}}
%\newcommand{\added}[1]{\textcolor{red}{#1}}
%\newcommand{\hide}[1]{~\textcolor{green}{[#1]}}
%\newcommand{\rewritten}[1]{\textcolor{blue}{#1}}
%\newcommand{\add}[1]{\textcolor{red}{#1}}

%\newcommand{\rewritten}[1]{#1}
%\newcommand{\delete}[1]{}
%\newcommand{\hide}[1]{}
%\newcommand{\added}[1]{#1}
\newcommand{\add}[1]{#1}

\section{Introduction}
During recent years, emerging short pulse high-intensity extreme ultraviolet (XUV) and X-ray sources such as laser-driven high-order harmonic generation (HHG)~\cite{FerrayJPB1988,AgostiniRPP2004,McPhersonJOSAB1987,CorkumPRL1993,SchaferPRL1993} and free-electron lasers (FELs)~\cite{Ackermann_2007, Emma_2010, Ishikawa_2012, McNeil_2010} have opened up new fields of science.~They made it possible to study ultrafast dynamics induced and probed with wavelengths in the XUV/X-ray regimes on femtosecond (FELs)~\cite{Glownia_2010} and even attosecond (HHG)~\cite{Lepine_2014,KrauszRMP2009} time scales.~In addition, the~high intensities and short pulse durations enable the study of hitherto inaccessible ionization processes~\cite{SorokinPRA2007, Berrah_2010, Bucksbaum_2011, Fang_2012}, as well as single shot imaging of macromolecular complexes~\cite{Gallagher-Jones_2014, Seibert_2011} by circumventing the resolution limit set by radiation damage for long exposures~\cite{Neutze_2000}.~However, compared to traditional laser or synchrotron experiments, experiments using short pulse high-intensity XUV and X-ray sources typically have to deal with very large single-shot signals ($\sim$1000 events/shot), often at relatively low repetition rates ($\sim$10--100 Hz), which~calls for adapted detection schemes.

In experiments employing photo-induced ionization or fragmentation, the~information about the interaction process lies in the energy and angular distribution of the emitted photo-electrons and -ions%please confirm this hyphen, if keep it? - It is a supended hyphen, and "photo-electrons" should be closed up to "photoelectrons" (no hyphen).
. One way of obtaining this information is by means of the application of a reaction microscope (REMI)~\cite{Ullrich03rpp}, where the initial three-dimensional momentum of particles is deduced from measurements of their impact coordinates on the detector, as well as from the flight time, acquired by means of, e.g., delay-line detectors. 
REMI gives access to complete, correlated, 3D velocity information of all particles detected in coincidence, as long as the count rates are sufficiently low ($<$1~event/shot) to avoid detecting fragments from more than one target atom or molecule on the same shot.

With high-intensity sources, the~repetition rates are typically low, while the single-shot count rates can be very high, which~in many cases makes it difficult to use techniques relying on detecting coincidences. Another approach is to use the so-called velocity map imaging (VMI) technique~\cite{EppinkRSI1997}, which~uses an extraction field configuration that makes the impact coordinates on the detector independent of the location of the ionization event within the interaction volume, as well as of the momentum along the detector axis. This allows for the use of a micro-channel plate (MCP) and a~phosphor screen where the impact of a large number of particles can be accumulated on every shot. Under the condition of cylindrical symmetry of the ionization process, the~initial three-dimensional momentum distribution of the particles can be recovered from the measured two-dimensional projection using numerical inversion procedures~\cite{Montgomery_Smith_1988, VrakkingRSI2001}. 

\textls[-10]{There are several demonstrations of using VMI, under low count rate conditions, for~photoelectron-photoion coincidence spectroscopy~(PEPICO)~\cite{Eland_1978}, for~example using electron VMI combined with ion time-of-flight (TOF), ion VMI together with electron spectroscopy and electron VMI with ion~VMI~\mbox{\cite{Bodi_2012,Sztaray_2017,Garcia_2009,OKeeffe_2011,Garcia_2005}}.} 

An obvious drawback of VMI in high count rate conditions is that one cannot rely on coincidence detection for extracting information about different electrons or ions coming from the same target molecule.~An elegant way to overcome this lack of correlated information in VMI, without~sacrificing the high count rates, is to use covariance mapping~\cite{FRASINSKI_1989}.~Briefly, for~any two \textls[-5]{variables, \mbox{$\mathbf{X}=\left[X_1,X_2, \dots, X_N \right]$} and $\mathbf{Y}=\left[Y_1,Y_2, \dots, Y_N \right]$, sampled synchronously in a repetitive measurement,} one can calculate the covariance, which~is a measure of how well correlated the variations of the two variables are.~Covariance mapping has successfully been applied in several laser and FEL experiments for different detection schemes, demonstrating ion-ion, electron-electron and~ion-electron covariance mapping~\cite{Frasinski_1992,Frasinski_2016,Kornilov_2013,Boguslavskiy_2012}.~There are also a few examples where not only the mass or energy, but~also the momentum of the particles has been studied through so-called covariance imaging~\cite{Zhu_1997,Hansen_2012,Slater_2015,Pickering_2016}. Recently, a double-sided electron-ion momentum imaging spectrometer has been installed at the Laser Applications in Material Processing (LAMP)%define if appropriate
 end-station at the Linac Coherent Light Source (LCLS). It features a complex lens design that allows operation either as a REMI or a VMI spectrometer, and~for the simultaneous detection of scattered XUV/X-ray photons on a pn charge-coupled device (pnCCD)%define if appropriate
 photon detector~\cite{Osipov_2018}. While offering high flexibility, such devices by necessity compromise the imaging capabilities to some extent, due to the adapted electrode design, as well as to the absence of additional shielding of the interaction region from external fields. 

Here, we present the design and performance of a double VMI spectrometer (VMIS) for covariance imaging of electrons and positively-charged ions, optimized for experiments using high-intensity XUV and X-ray sources. With minimal compromising of the imaging capabilities, the~instrument is versatile and allows for combining photo-electron and -ion detection modes, such~as ion TOF, ion VMI and~electron VMI, in different ways, depending on the required information and the process under study. The performance of the different detection modes is estimated with simulations and tested in experiments using an intense HHG source, with promising first results using the ion TOF signal for species selection in ion VMI and extraction of channel-specific information from electron VMI data.

In Section~\ref{sec:apparatus}, we describe the design of the apparatus, discuss technical requirements and the motivation of important design details such as the electrode geometry and the mounting, which allow floating voltages. Further, we describe the simulation procedure and present data on the estimated performance in the different operation modes.~In Section~\ref{sec:experiments}, we show results from measurements with the spectrometer at the high-intensity XUV beamline at the Lund Attosecond Science Center. Finally,~we~conclude in Section~\ref{sec:conclusion}.

\section{Apparatus}
\label{sec:apparatus}

\begin{figure}[H]
\centering
\includegraphics[width=0.5\textwidth]{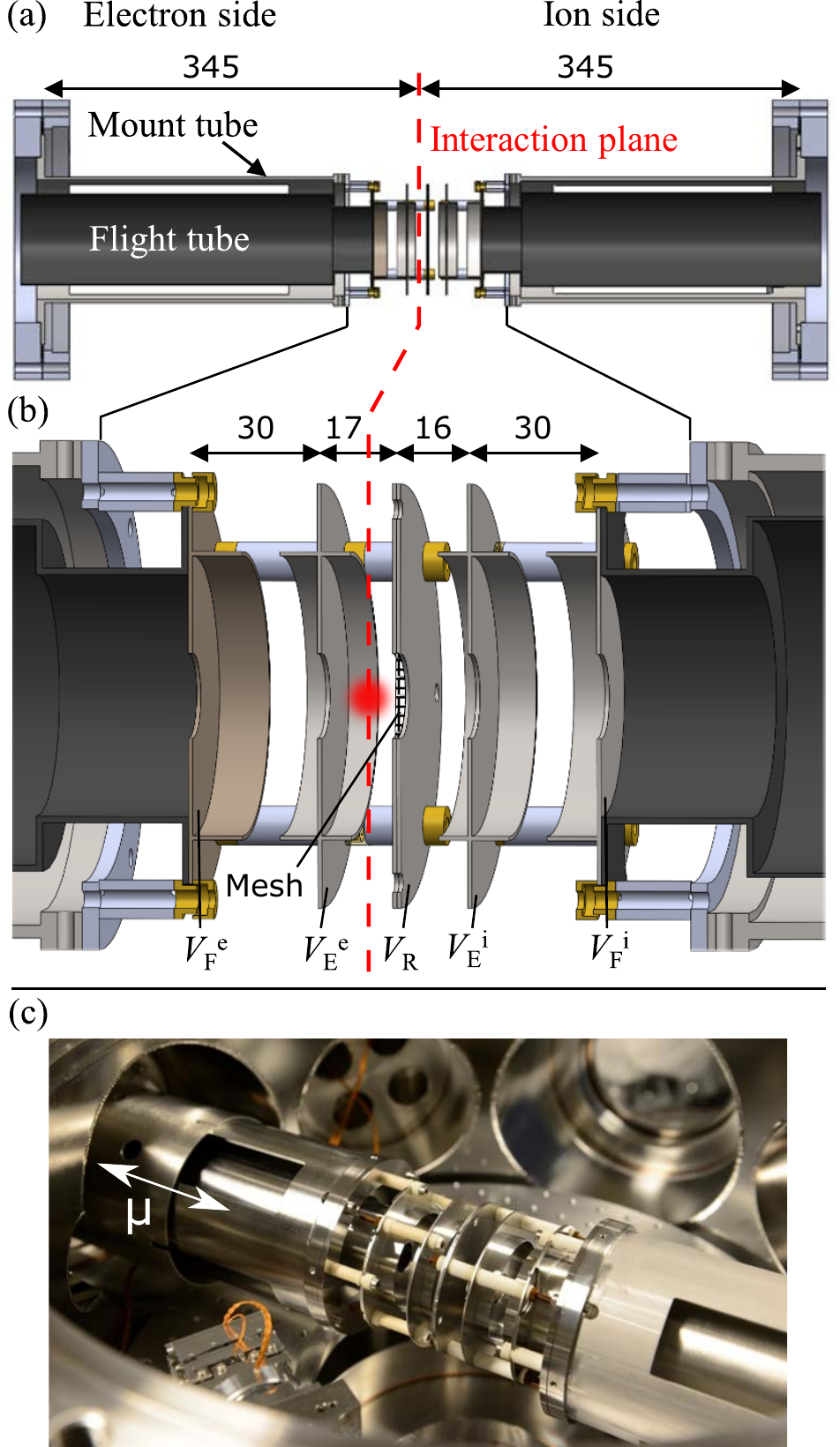}
\caption{Drawing of the double velocity map imaging spectrometer (VMIS) (\textbf{a}) and an expanded view of the electrode package (\textbf{b}); dimensions are in mm.~The indicated voltages are applied to the electron side flight tube ($V_\mathrm{F}^\mathrm{e}$), the~electron side extractor ($V_\mathrm{E}^\mathrm{e}$), the~repeller ($V_\mathrm{R}$), the~ion side extractor ($V_\mathrm{E}^\mathrm{i}$) and the ion side flight tube ($V_\mathrm{F}^\mathrm{i}$). A photo of the spectrometer mounted in the experimental chamber is shown in (\textbf{c}). The~sliding $\mu$-metal shield is marked by the letter $\mu$.}
\label{fig:DVMIS_design}
\end{figure}

The design of the double VMIS is an extension of the standard VMIS suggested by Eppink and Parker in 1997~\cite{EppinkRSI1997}. The standard VMIS, corresponding to the left side in Figure~\ref{fig:DVMIS_design}a, consists of a~repeller plate, an open extractor plate and~a field-free flight tube. In the simplest case, where the flight tube is grounded ($V_\mathrm{F}^e=0$), the~ratio between the extractor voltage ($V_\mathrm{E}^e$) and the repeller voltage ($V_\mathrm{R}$) defines the imaging mode of the VMIS. In the general case, the~ratio can be written as:
\begin{equation}
\eta=\frac{V_\mathrm{E}^\mathrm{e}-V_\mathrm{F}^\mathrm{e}}{V_\mathrm{R}-V_\mathrm{F}^\mathrm{e}}.
\label{eq:onlyone}
\end{equation}
\unskip

By changing $\eta$, the~curvature of the field in the interaction region is changed, allowing for different imaging modes.~Velocity map imaging mode is usually achieved for $\eta \approx 0.75$, with the exact value depending on the specific design geometry~\cite{EppinkRSI1997}.~In this mode, electrons (or ions) with the same momentum in the plane parallel to the detection plane will be projected onto the same spot on the detector even if they are generated at different positions in the interaction region. 
 
The full three-dimensional momentum distribution can be obtained through inversion procedures, as long as there is an axis of cylindrical symmetry in the ionization process~\cite{Montgomery_Smith_1988, VrakkingRSI2001}. Another mode of operation, often referred to as the spatial imaging mode, is achieved for $\eta \approx 1$~\cite{Johnsson_2010}. The particle impact position on the detector is then most sensitive to where the charged particle is created, and therefore, the interaction region can be imaged. This is a useful mode for alignment purposes, providing a means of ensuring precise positioning of the XUV light focal spot, as well as good overlap between the laser beam and the molecular beam~\cite{Johnsson_2010}. Moreover, by choosing $\eta $ between 0.78 and 0.8, a Wiley--McLaren ion time-of-flight mode can be achieved~\cite{Wiley_1955}.

\subsection{Design}
The design of the double VMIS was made with the following goals: First, we wanted to add the capability to detect positively-charged ions without compromising the resolution on the electron side as compared to a standard VMIS. 
Second, as molecules have a typical ionization potential of 7--15~eV and, with HHG, the~typical photon energies are up to 100~eV, the~spectrometer should be able to focus electrons with energies up to $\approx$90~eV and~ions with kinetic energies up to $\approx$10~eV, typical for, e.g., Coulomb explosion.
Third, the~spectrometer had to be compatible with the existing experimental chamber of the high-intensity XUV beamline at the Lund Attosecond Science Center containing the all-reflective short focal length XUV focusing optics~\cite{ManschwetusPRA2016,Coudert-Alteirac_2017}.

Extending the single-sided spectrometer to be able to record positively-charged ions and electrons simultaneously was the aim of this work. To that end, the~design of the electron side, adapted from~[\citenum{RoscaPruna2001}], was~retained and the repeller electrode replaced by an electrode with a mesh, as shown in Figure~\ref{fig:DVMIS_design}b.~On the ion side, an open extractor electrode and a flight tube, similar to that on the electron side, were added.~There are two advantages with this choice of design.~First, we are not changing the imaging conditions on the electron side, and second, once the optimum voltages have been found for the electrons, the~voltages on the ion side can be tuned independently without affecting the electron image quality.~The drawbacks are that the ions are to some extent scattered by the mesh, which~has a transmission of $\sim$80\%, and~that they pass through a drift region before they pass the repeller and are focused by the fields on the ion side. This affects the energy resolution that can be achieved for the ion imaging, which~will be discussed in Section~\ref{sec:simulations}. 

The maximum electron or ion energy is restricted by the fact that the electrons or ions move away from the detector axis due to their initial velocity perpendicular to it. To increase the maximum detectable energy, one needs to either decrease the flight tube length, increase the flight tube and detector diameters or~use higher acceleration voltages. In our case, the~maximum detector size and the flight tube diameters were set by the existing chamber to $\sim$100~mm, leading to the choice of a standard MCP detector with a diameter of 75~mm, mounted on a CF160 flange. Similarly, the~minimum flight tube lengths were set by the flange-to-flange distance (690 mm) of the existing experimental chamber to 345~mm. In principle, shorter flight tube lengths could have been achieved by a design where the MCPs are mounted inside the vacuum chamber, but~this would have led to a more complicated design and a deteriorated TOF resolution, and~was thus avoided. 

With restrictions imposed on the detector size and flight tube length, the~ratio between the maximum detectable kinetic energy and the acceleration voltage scales approximately as $\left(r/L\right)^2$, where $r$ is the detector radius and $L$ is the flight tube length. Under the chosen conditions, this means that particles with kinetic energies up to $\approx$1\% of the acceleration voltage can be detected, meaning that a total voltage difference of almost 10~kV between the front of the electron MCP and the front of the ion MCP is required to detect electrons and ions with 90~eV and 10~eV of kinetic energy, respectively. 

The design of the assembly was made in such a way that both detectors, electrodes and flight tubes could be supplied with floating voltages of $\pm 10$~kV. While the front of the MCPs are not electrically connected to the flight tubes, for~the following discussion and simulations, we assume that the same voltage is applied to the front of the MCP and the flight tube.

The resulting design of the double VMIS is shown in Figure~\ref{fig:DVMIS_design} together with the relevant dimensions.
The spectrometer is mounted horizontally, and~the electron and ion sides are separable between the repeller electrode and the ion extractor electrode to facilitate mounting of the spectrometer on the two opposite CF200 flanges of the experimental chamber. The mounting flanges are CF200 to CF160 zero-length reducers, allowing for convenient mounting of the detectors after the spectrometer has been installed. In order to allow for floating voltages on the flight tubes, ceramic spacers are used between these and the supporting grounded mount tubes, which are made of stainless steel. While~the ion flight tube is made of stainless steel, the~electron flight tube is made of $\mu$-metal to provide shielding from external magnetic fields. 
In addition, a sliding $\mu$-metal cylinder with holes for the laser beam and the molecular beam, visible on the far left in Figure~\ref{fig:DVMIS_design}c, can be moved in to shield also the electrode assembly.
The flight tubes have a front brim, which is supported by four rods via ceramic spacers (white parts in Panel (b)) and bushings (yellow parts in Panel (b)).
These brims also hold the electrodes, which can be conveniently reached and exchanged from the top of the chamber without unmounting the flight tube assemblies. 
The electrodes are mounted with ceramic bushings (yellow parts in Panel (b)) on rods made out of Vespel\textsuperscript{\textregistered}, separated by ceramic spacers (white parts in Panel (b)). 
No electrode or grounded parts are closer than 10~mm apart to allow for large voltage differences without the risk of arcing. 

The final design offers a large versatility by allowing for voltages of $\pm10$~kV to be applied to any of the spectrometer components (electrodes or flight tubes), and~the choice of a mesh in the repeller electrode allows for independent tuning of the ion sides without affecting the electron imaging. The~physical design makes it easy to mount and dismount the whole or parts of the spectrometer, and while being dimensioned for use in the existing experimental chamber, the~assembly can be easily transferred also to chambers with other geometries.

\subsection{Operation Modes}

The spectrometer can be operated in different photo-electron and -ion detection modes. A selection of spectroscopically-relevant modes is listed below:
\begin{enumerate}
\item High resolution ion modes:
\begin{enumerate}
\item Ion TOF,
\item Ion VMI.
\end{enumerate}
\item High resolution electron modes:
\begin{enumerate}
\item Electron VMI, ion TOF,
\item Electron VMI, ion VMI.
\end{enumerate}
\end{enumerate}

In the high resolution ion modes, the~voltages are set to accelerate and detect ions on the electron side, as a standard single-side VMI, either in TOF mode (1(a)) %please confirm if change the “[” “]” to “(” “)”. - This is okay with us, if it suits the style of the journal better
under Wiley-McLaren conditions~\cite{Wiley_1955} or~in VMI mode (1(b)).~With this setting, the~mesh in the repeller electrode does not impose any restriction on the achievable mass or energy resolution, since the ion side of the spectrometer is not used (see Figure~\ref{fig:DVMIS_design}).~In the high resolution electron modes, the~electron side of the spectrometer is configured for optimum VMI conditions for electrons, while the ion side voltages are set either for ion TOF (2(a)) or ion VMI (2(b)). The different modes are explored in Section~\ref{sec:simulations}.

\subsection{Simulations}
\label{sec:simulations}

To evaluate the expected performance of the spectrometer, simulations were carried out using SIMION~\cite{Simion} for the different operation modes listed above.~An interaction region, defined by the overlap between the laser beam and the molecular beam, with a conservatively chosen size of \mbox{$(100~\times100\times2000)$ $\upmu$m$^3$} was used. For the VMI calculations, electron kinetic energies were chosen in steps of 5 eV from 10--90 eV and ion kinetic energies in steps of 1 eV from 2 to 10 eV. For each operation mode, we found optimum voltages, defined as the voltages for which the best resolution is obtained for electrons with a kinetic energy of 30~eV and singly-charged ions with a kinetic energy of 6~eV and a mass of $m=100$~u. %please check the convention; define if appropriate - changed according to convention, also in all other occurences including figure 6(d)
 The kinetic energies were chosen based on the typical energies of the photoelectrons and photoions of interest using high-order harmonics generated in argon to ionize molecules with ionization potentials in the 10--20-eV range, and~the mass was chosen because it is typical for systems that are of current scientific interest, like small polycyclic aromatic hydrocarbons.

To find the optimum voltages for the VMI modes, a simplified simulation procedure was used, in which the trajectories of 27 particles arranged on a $3\times3\times3$ grid covering the interaction region and with their velocity component perpendicular to the detector axis were calculated and the energy or mass resolution calculated from the particle impact position distribution on the detector. For the optimum voltages found (summarized in Table~\ref{tab:voltages}), a procedure based on Monte Carlo sampling was used~\cite{GhafurRSI2009}, resulting in more realistic resolution estimates. The method consisted of launching a large number of electrons or ions ($\sim$10$^6$) with a random starting position chosen from a Gaussian distribution over the interaction volume and with a random, isotropically-distributed 3D initial momentum.~The particle impacts on the detector were then sampled to generate a simulated detector image or TOF trace, which~was treated similarly to experimental data in order to extract the resolution.~The simulated images were inverted using an iterative algorithm~\cite{VrakkingRSI2001} to retrieve the 3D momentum distribution from the 2D image.
From the 3D momentum distribution, the energy spectrum was calculated, and the energy resolution, $\Delta E/E$, was~estimated by obtaining $\Delta E$ from the full width at half maximum of a Gaussian fit to the peaks in the spectrum.
For the TOF simulations, only the Monte Carlo method was used, and the mass resolution, $m/\Delta m$, was~calculated by obtaining $\Delta m$ from the full width at half maximum of a Gaussian fit to the peaks in the mass spectrum.
They were performed both for ions with zero kinetic energy and for ions with an initial velocity in the direction of the molecular beam. The latter represents the more realistic scenario of target molecules with a mass of 100~u that are brought into the interaction region by a carrier gas, e.g., helium. This carrier gas typically travels at a~speed of 1000 m/s, which~corresponds to a kinetic energy of 520~meV for molecules with a mass of 100~u.
\begin{table}[htbp]
	\centering
		\begin{tabular}{lcccccc}
		\toprule
		\textbf{Operation Mode} & \boldmath$\eta$ \textbf{(Equation~(\ref{eq:onlyone}))} & \boldmath{$V_\text{F}^\text{e}$} \textbf{(kV)} & \boldmath{$V_\text{E}^\text{e}$} \textbf{(kV)} & \boldmath{$V_\text{R}$} \textbf{(kV)} & \boldmath{$V_\text{E}^\text{i}$} \textbf{(kV)} & \boldmath{$V_\text{F}^\text{i}$} \textbf{(kV)} \\ 
		\midrule
		1(a) Ion TOF & 0.76--0.78 & 0 & 2.275--2{.}354 & 3.000 & not used & not used \\ 
		1(b) Ion VMI & 0.76 & 0 & 2.275 & 3.000 & not used & not used \\
		2(a) Electron VMI + ion TOF & 0.76 & 5.000 & $-$5.775 & $-$9.200 & $-$10.000 & $-$10.000 \\ 
		2(b) Electron VMI + ion VMI & 0.76 & 5.000 & $-$5.775 & $-$9.200 & $-$8.802 & $-$10.000 \\
		\bottomrule		
		\end{tabular}
	\caption{Simulated optimum voltages for the different operation modes.~For Mode 1(a), the~exact optimum voltage $V_\text{E}^\text{e}$ depends on the initial velocity of the ions. }
	\label{tab:voltages}
\end{table}

\subsubsection{High Resolution Ion Modes}
\begin{figure}[H]
\centering
\begin{tabular}{rl}
%\multicolumn{2}{c}{
\includegraphics[width=0.5\linewidth]{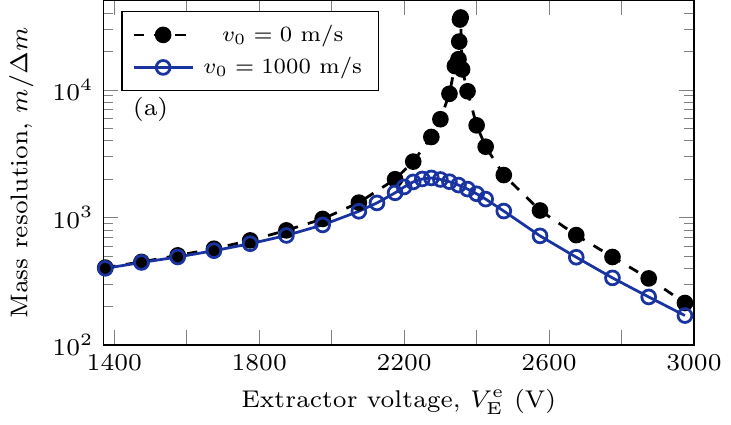}%}\\
%\multicolumn{2}{c}{
& \includegraphics[width=0.45\linewidth]{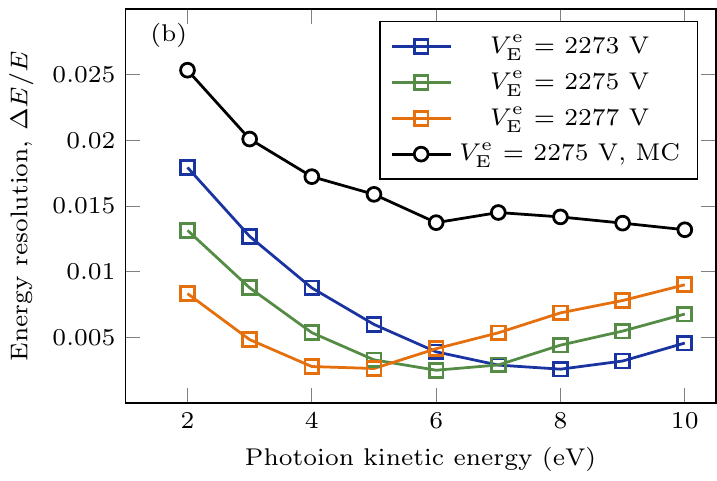}\\%}\\
\includegraphics[width=0.3\linewidth]{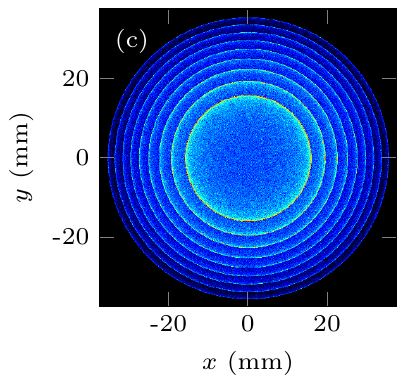}
&\includegraphics[width=0.414\linewidth]{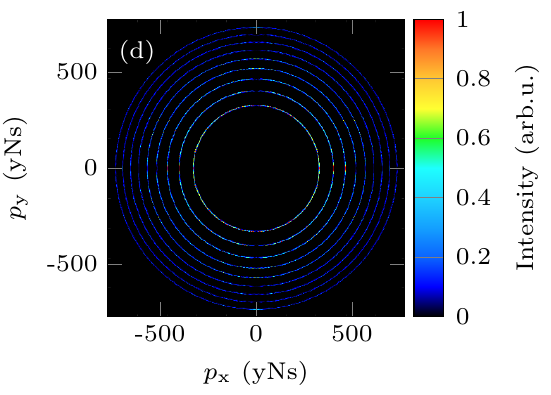}\\
\multicolumn{2}{c}{\includegraphics[width=0.48\linewidth]{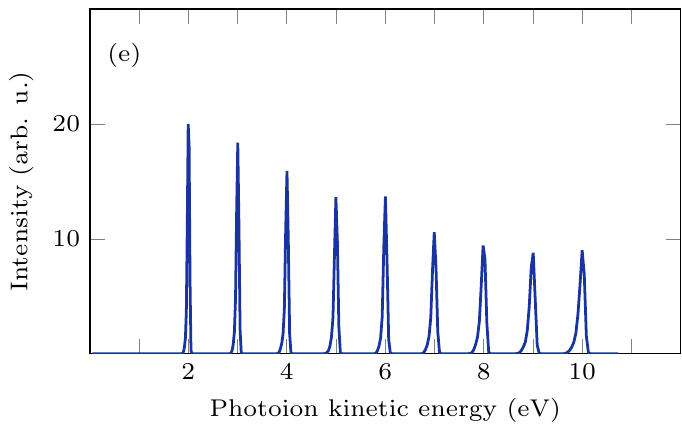}}\\
\end{tabular}
%
%  \begin{subfigure}{}
%   \includegraphics[]{}
%  \end{subfigure}
%	\begin{subfigure}{}
%   \includegraphics[]{}
%   \label{fig:DSionsEside}
%  \end{subfigure}
%		
%   \hspace*{14mm}
%		\begin{subfigure}{}
%   \includegraphics[]{}
%   \label{fig:VMI_ions_eside}
%		\end{subfigure}
%   \hspace*{-2mm}
%		\begin{subfigure}{}
%   \includegraphics[]{}
%   \label{fig:VMI_ions_eside_inv}
%		\end{subfigure}
%				
%		\hspace*{2.9mm}\begin{subfigure}{}
%   \includegraphics[]{}
%   \label{fig:DSions_PIS}
%  \end{subfigure}
\caption{Simulation results for the high resolution ion modes. (\textbf{a}) shows the resulting mass resolution at mass $m=100$~u %define if appropriate
 as a function of extractor voltage, $V_\mathrm{E}^\mathrm{e}$, for~ions with zero initial kinetic energy (black) and ions with an initial velocity of $v_0=1000$~m/s in the direction of the molecular beam (blue). (\textbf{b}) shows the resulting VMI energy resolution for three different extractor voltages ($V_\mathrm{E}^\mathrm{e}$) using the 27-particle model (blue, green and orange lines) and the Monte Carlo model for the optimum setting (black line). (\textbf{c},\textbf{d}) show the simulated detector image and the 3D momentum distribution after inversion, for~the optimum voltages. (\textbf{e}) shows the kinetic energy spectrum calculated from the 3D momentum distribution.}
\label{fig:mode1} % We appreciate your effort to save space here, but we think it is not aesthetically pleasing in the current version with panels a nd b in the same row especially in contrast to the second row of panels c and d. The same applies in principle to Figure 4.
\end{figure}

For the high resolution ion modes, ions are detected on the electron side of the spectrometer, and~for the simulations, the choice was made to ground the front of the MCP and the flight tube ($V_\mathrm{F}^\mathrm{e}=0$~kV). The repeller voltage was set to $V_\mathrm{R}=3$~kV. Figure~\ref{fig:mode1}a shows the resulting mass resolution in Operation Mode 1(a) for extractor voltages from 1400--3000~V. 
For zero kinetic energy ions, the~optimum extractor voltage was found to be $V_\mathrm{E}^\mathrm{e}=2.354$~kV ($\eta = 0.78$, black curve). The resolution of $m/\Delta m=$ 35,000 is clearly over-estimated, and~for the more realistic case of ions with an initial drift velocity, the best resolution of $m/\Delta m=2000$ was found for a voltage of $V_\mathrm{E}^\mathrm{e}=2.275$~kV ($\eta = 0.76$, blue~curve).

Figure~\ref{fig:mode1}b shows the resulting energy resolution for ion VMI in Operation Mode 1(b) for three different settings for the extractor voltage, $V_\mathrm{E}^\mathrm{e}$, calculated using the 27-particle model (blue, green and orange lines). The optimum extractor voltage, for~which the best resolution was achieved for a~photoion kinetic energy of 6~eV, was~found to be $V_\mathrm{E}^\mathrm{e}=2.275$~kV ($\eta = 0.76$), and for this setting, the resulting energy resolution from the Monte Carlo simulation is shown (black line), indicating an energy resolution better than $1.5\%$ for kinetic energies between 6 and 10~eV. In Figure~\ref{fig:mode1}c,d, the corresponding simulated detector image and the 3D momentum distribution after inversion are shown, respectively. Figure~\ref{fig:mode1}e shows the kinetic energy spectrum calculated from the 3D momentum distribution.

\subsubsection{High Resolution Electron Modes}

For the high resolution electron modes, electrons and ions are detected on their respective sides of the spectrometer. 
To maximize the possible total voltage difference over the spectrometer, the~voltage on the electron side flight tube (and thus, the front of the electron MCP) was set to $V_\mathrm{F}^\mathrm{e}=5$~kV, so~that the phosphor screen could still be operated at its full voltage bias of 5~kV relative to the front of the MCP without exceeding the limit of the electrical feedthroughs.
The repeller voltage was set to $V_\mathrm{R}=-9.2$~kV to allow for detection of electrons with kinetic energies up to 90~eV, while allowing for extraction and imaging of ions with kinetic energies up to 10~eV.

Figure~\ref{fig:mode2}a shows the resulting energy resolution for electron VMI for three different settings for the extractor voltage, $V_\mathrm{E}^\mathrm{e}$, calculated using the 27-particle model (blue, green and orange lines). The~optimum extractor voltage, for~which the best resolution was achieved for a photoelectron energy of 30~eV, was~found to be $V_\mathrm{E}^\mathrm{e}=-5.775$~kV ($\eta=0.76$), and for this setting, the resulting energy resolution from the Monte Carlo simulation is shown (black line), indicating an energy resolution better than $2\%$ for photoelectron energies between 30 and 85~eV. In Figure~\ref{fig:mode2}b,c, the corresponding simulated detector image and the 3D momentum distribution after inversion are shown, respectively. Figure~\ref{fig:mode2}d shows the photoelectron energy spectrum calculated from the 3D momentum distribution.
\begin{figure}[H]
\centering
\begin{tabular}{rl}
\multicolumn{2}{c}{\includegraphics[width=0.5\linewidth]{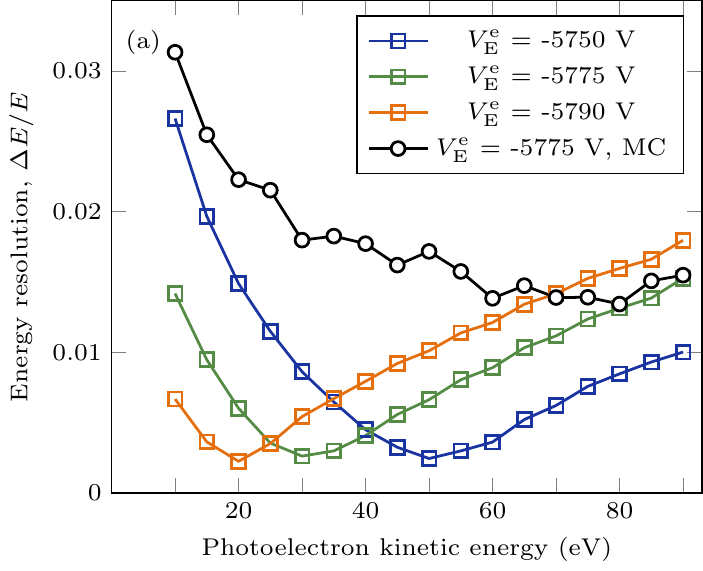}}\\
%\multicolumn{2}{c}{
\includegraphics[width=0.3\linewidth]{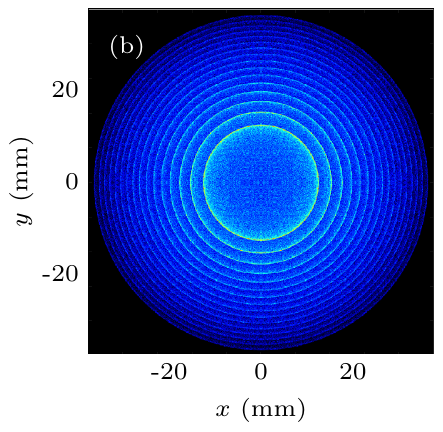}
& \includegraphics[width=0.396\linewidth]{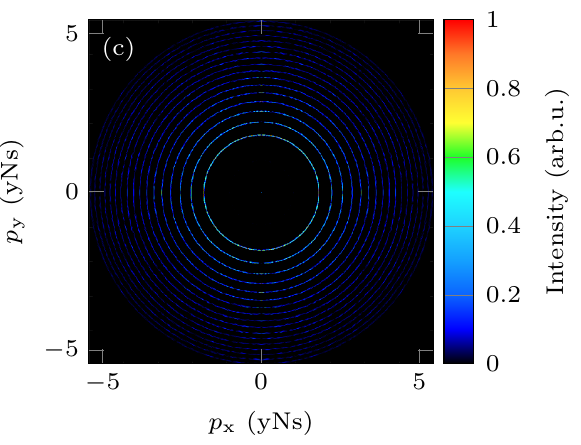}\\
\multicolumn{2}{c}{\includegraphics[width=0.52\linewidth]{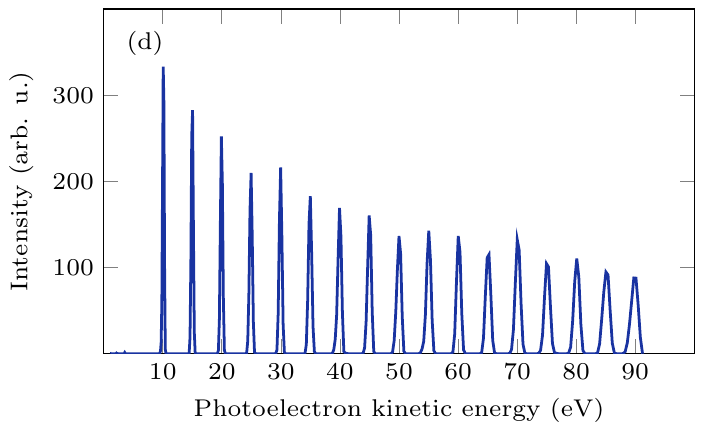}}\\
\end{tabular}
%		\begin{subfigure}{}
%   \includegraphics[]{}
%   \label{fig:DVMIS_zoom}
%  \end{subfigure}
%		
%   \hspace*{16mm}
%		\begin{subfigure}{}   \includegraphics[]{}
%   \label{fig:DVMIS}
%		\end{subfigure}
%   \hspace*{-2mm}    
%		\begin{subfigure}{}   \includegraphics[]{}
%   \label{fig:DVMIS}
%		\end{subfigure}
%				
%		\begin{subfigure}{}
%   \includegraphics[]{}
%   \label{fig:DVMIS}
%  \end{subfigure}
\caption{Simulation results for high resolution electron VMI mode. (\textbf{a}) shows the resulting energy resolution for three different extractor voltages ($V_\mathrm{E}^\mathrm{e}$) using the 27-particle model (blue, green and orange lines) and the Monte Carlo model for the optimum setting (black line). (\textbf{b},\textbf{c}) show the simulated detector image and the 3D momentum distribution after inversion, for~the optimum voltages. (\textbf{d}) shows the photoelectron energy spectrum calculated from the 3D momentum distribution.}
   \label{fig:mode2}
\end{figure}

With the electron side conditions optimized for electron VMI and~the voltage on the ion side flight tube set to $V_\mathrm{F}^\mathrm{i}=-10$~kV to detect ions with kinetic energies up to 10 eV, the~ion side can be tuned either for TOF (Operation Mode 2(a)) or VMI (Operation Mode 2(b)) by varying the ion extractor electrode voltage, $V_\mathrm{E}^\mathrm{i}$. 

Figure~\ref{fig:mode2ab}a shows the resulting mass resolution in Operation Mode 2(a) for ions with an initial velocity of 1000~m/s in the direction of the molecular beam, for~different extractor voltages, $V_\mathrm{E}^\mathrm{i}$. 
It~is clear that it is not possible to reach Wiley--McLaren conditions within the voltage limits of the setup, due to the initial drift region between the interaction point and the mesh through which the ions have to travel. In addition, the~best mass resolution of $m/\Delta m=110$, achieved for $V_\mathrm{E}^\mathrm{i}=-10$~kV, is~a~factor of 18 worse than that achieved in Operation Mode 1(a). 
In fact, the~mass resolution is in this case limited by the influence of the drift region and not by the initial velocity of the ions, and~the mass resolution for zero kinetic energy ions (not shown) coincides with the mass resolution shown in Figure~\ref{fig:mode2ab}a. Figure~\ref{fig:mode2ab}b shows the resulting energy resolution for ion VMI in Operation Mode 2(b) for three different settings for the extractor voltage, $V_\mathrm{E}^\mathrm{i}$, calculated using the 27-particle model (blue, green~and orange lines). The~optimum extractor voltage, for~which the best resolution was achieved for a kinetic energy of 6~eV, was~found to be $V_\mathrm{E}^\mathrm{i}=-8.802$~kV, and for this setting, the resulting energy resolution from the Monte Carlo simulation is shown (black line), indicating an energy resolution better than $2.5\%$ for kinetic energies between 3 and 10~eV. This is a small degradation in resolution as compared to the resolution obtained in Operation Mode 1(b) (shown by the dashed black line), but~it is still an~acceptable resolution for ions in most experiments. It is interesting to note that even in Operation Mode 2(b), for~$V_\mathrm{E}^\mathrm{i}=-8.802$~kV, the~mass resolution of the ion TOF is $m/\Delta m=100$, which~is acceptable unless isotope resolution of heavier fragments is required.
In Figure~\ref{fig:mode2ab}c,d, the corresponding simulated detector image and the 3D momentum distribution after inversion are shown, respectively. Figure~\ref{fig:mode2ab}e shows the kinetic energy spectrum calculated from the 3D momentum distribution.
\begin{figure}[H]
\centering
\begin{tabular}{rl}
%\multicolumn{2}{c}{
\includegraphics[width=0.48\linewidth]{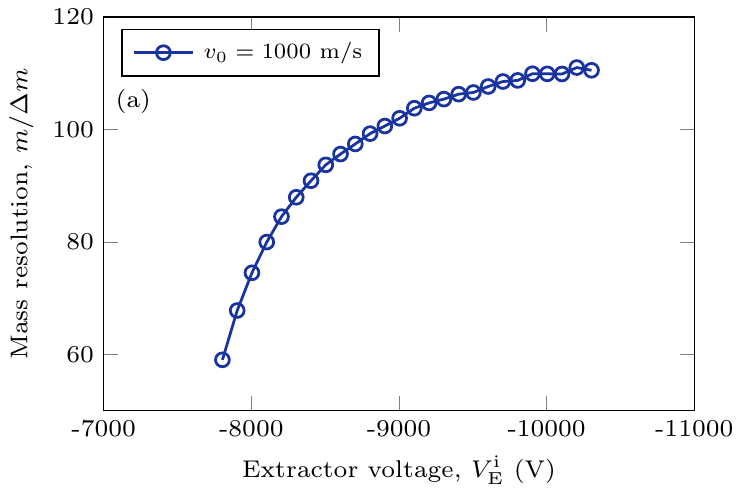}%}\\
%\multicolumn{2}{c}{
& \includegraphics[width=0.47\linewidth]{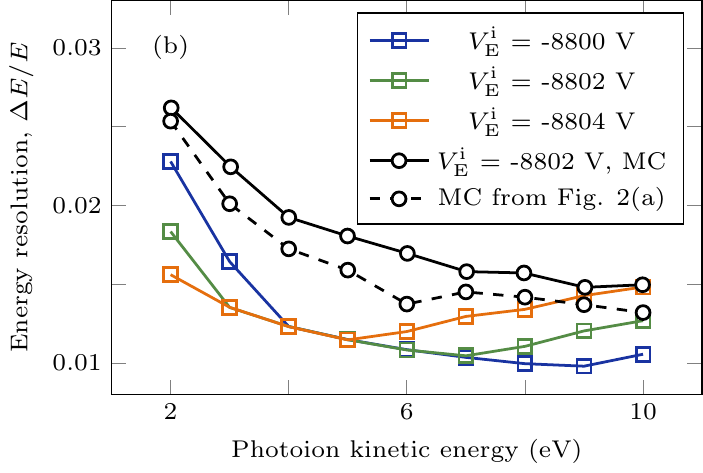}\\%}\\
\includegraphics[width=0.32\linewidth]{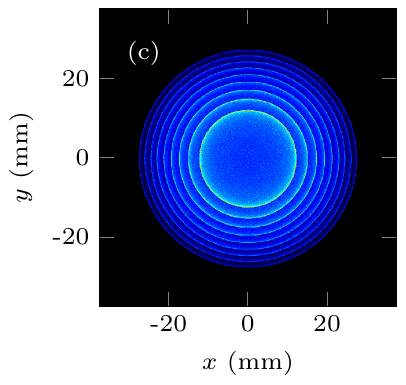}
&\includegraphics[width=0.44\linewidth]{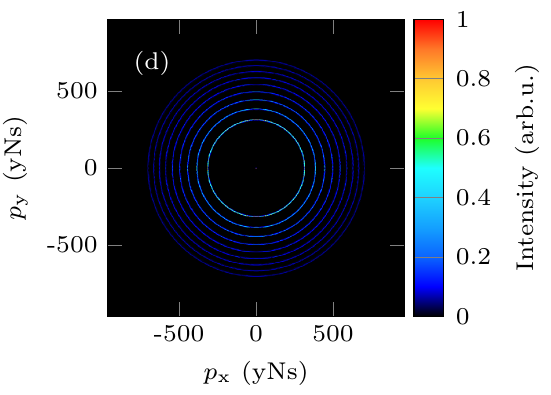}\\
\end{tabular}
\caption{{Cont.}}
\end{figure}

\begin{figure}[H]\ContinuedFloat
\centering
\begin{tabular}{rl}
\multicolumn{2}{c}{\includegraphics[width=0.48\linewidth]{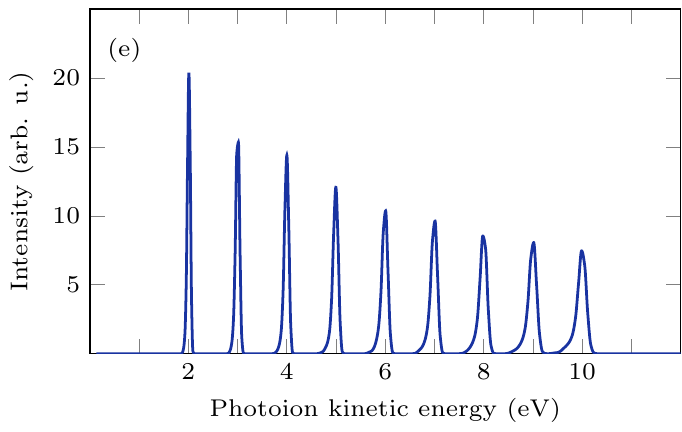}}\\
\end{tabular}
%\hspace*{5mm}
%		\begin{subfigure}{}
%   \includegraphics[]{}
%   \label{fig:DVMIS_zoom}
%  \end{subfigure}
%		
%\vspace*{-1mm}	
%		\begin{subfigure}{}
%   \includegraphics[]{}
%   \label{fig:DVMIS}
%		\end{subfigure}
%
%	\vspace*{-3mm}
%	\hspace*{14mm}
%		\begin{subfigure}{}
%   \includegraphics[]{}
%   \label{fig:VMI_ions_iside}
%		\end{subfigure}
%   \hspace*{-2mm}
%		\begin{subfigure}{}
%   \includegraphics[]{}
%   \label{fig:VMI_ions_iside_inv}
%		\end{subfigure}
%			
%	\vspace*{-2mm}	
%		\hspace*{2.9mm}\begin{subfigure}{}
%   \includegraphics[]{}
%   \label{fig:Ions_PIS_IS}
%  \end{subfigure}
\caption{Simulation results for ion detection when the electron side is optimized for electron VMI (Operation Modes 2(a) and 2(b)). (\textbf{a}) shows the resulting mass resolution as a function of extractor voltage, $V_\mathrm{E}^\mathrm{i}$, for~ions with an initial velocity of $v_0=1000$~m/s in the direction of the molecular beam. (\textbf{b}) shows the resulting VMI energy resolution for three different extractor voltages using the 27-particle model (blue, green and orange lines) and the Monte Carlo model for the optimum setting (solid black line). The dashed black line shows the resolution in Operation Mode 1(b) obtained by the Monte Carlo method, reproduced from Figure~\ref{fig:mode1}. (\textbf{c},\textbf{d}) show the simulated detector image and the 3D momentum distribution after inversion, for~the optimum voltages. (\textbf{e}) shows the kinetic energy spectrum calculated from the 3D momentum distribution.}
\label{fig:mode2ab} % We do not approve of the status of this figure. It is desirable to not seperate it onto two pages. We would appreciate if you could please change this accordingly. 
\end{figure}

\section{Experimental Results}

\label{sec:experiments}

\begin{figure}[H]\centering		 \includegraphics[width=0.5\columnwidth]{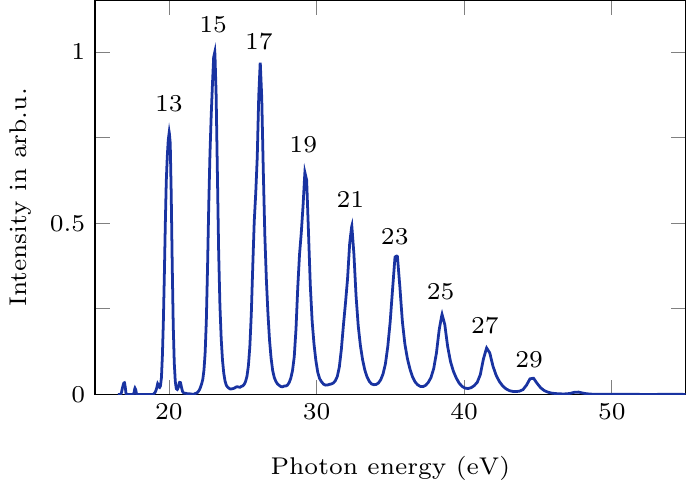}	
\caption{XUV spectrum recorded by the XUV spectrometer. The spectrum contains photons from harmonic order 13 (20.0 eV) up to harmonic order 29 (44.7 eV), indicated above the peaks.}
\label{fig:XUVspec}\end{figure}

To evaluate the performance of the instrument, measurements have been done at the high-intensity XUV beamline at the Lund Attosecond Science Center~\cite{ManschwetusPRA2016}, where high-flux XUV pulses generated through the HHG process were focused by reflective optics into the interaction region of the double VMIS. The target gas was introduced using a pulsed molecular beam from an Even--Lavie solenoid valve, able to produce high density pulses with durations as short as 10~$\upmu$s~\cite{Even_2014, Even_2015}. In Figure~\ref{fig:XUVspec}, the spectrum of the XUV pulses generated in argon and recorded by an XUV spectrometer is displayed. It~contains photons from harmonic order 13 (20.0 eV) up to harmonic order 29 (44.7 eV). 
Lower photon energy contributions and the remaining driving laser light are absorbed by a 200 nm-thick aluminum filter.
The XUV light is redirected to the XUV spectrometer by a rotatable gold mirror prior to the focusing into the interaction region of the double VMIS, and~thus, the spectrum cannot be recorded during measurements with the double VMIS. {The displayed XUV spectrum has been corrected for the spectral responses of the employed grating and MCP detector. In the further analysis, the corrected spectrum is used.}

\begin{figure}[H]
\centering
\begin{tabular}{lll}
\includegraphics[width=0.3\linewidth]{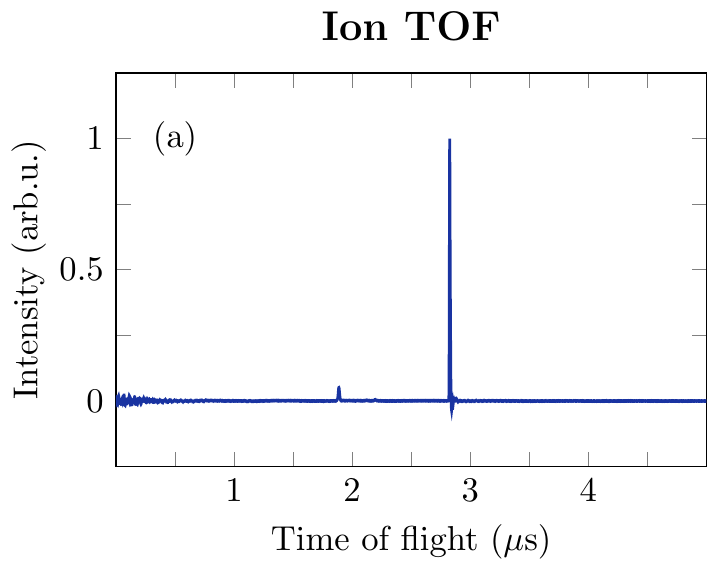}
&\includegraphics[width=0.3\linewidth]{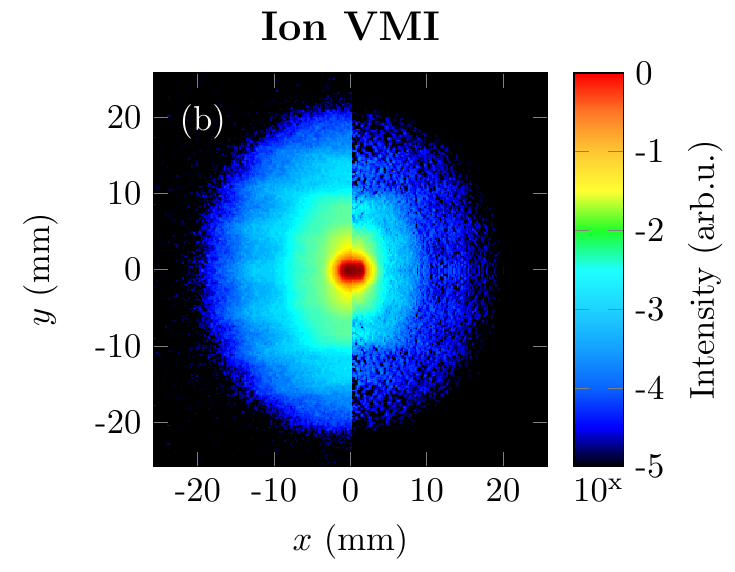}
&\includegraphics[width=0.33\linewidth]{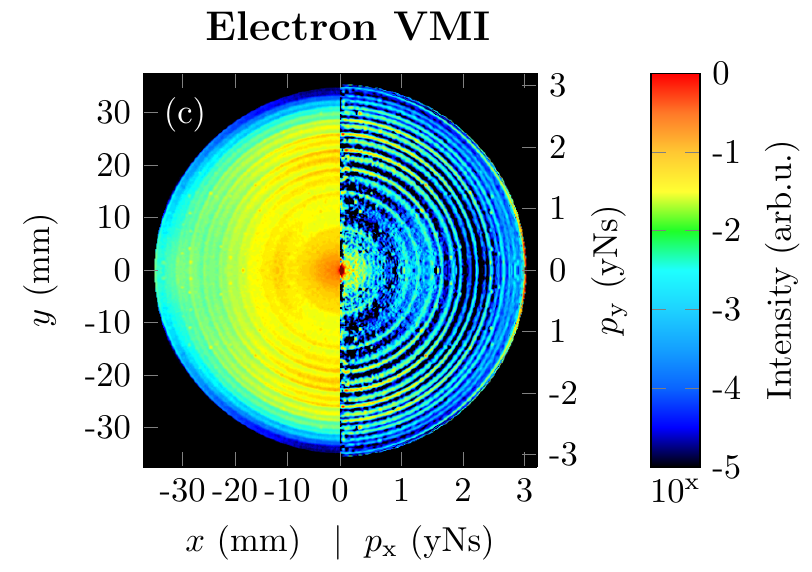}\\
\includegraphics[width=0.3\linewidth]{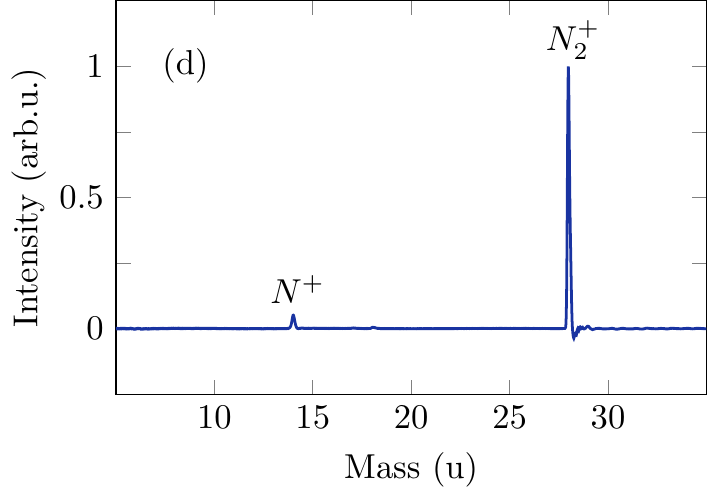}
&\includegraphics[width=0.3\linewidth]{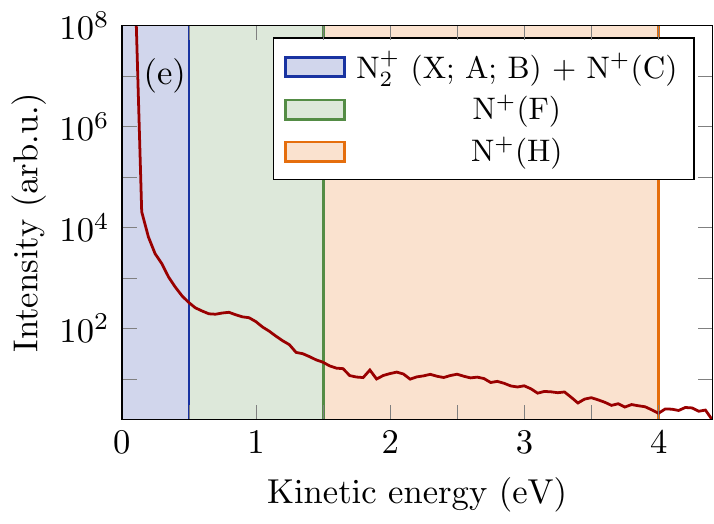}
&\includegraphics[width=0.31\linewidth]{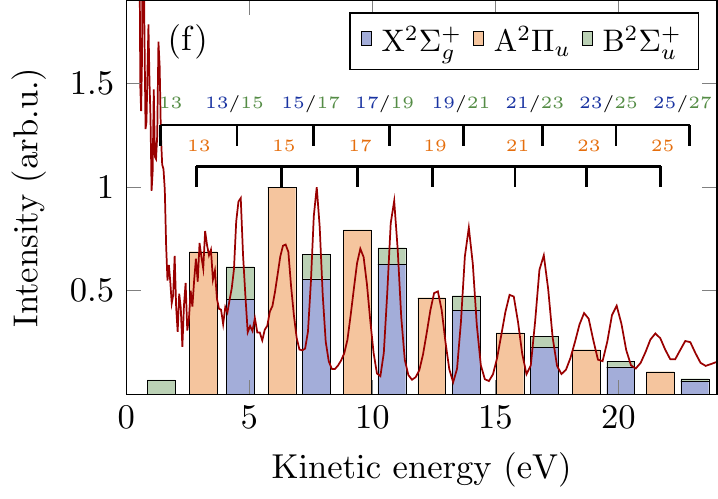}\\
\includegraphics[width=0.3\linewidth]{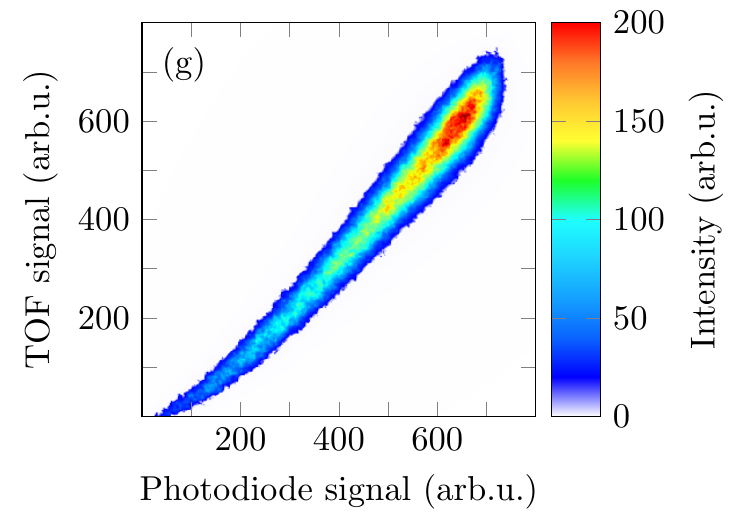}
&\includegraphics[width=0.3\linewidth]{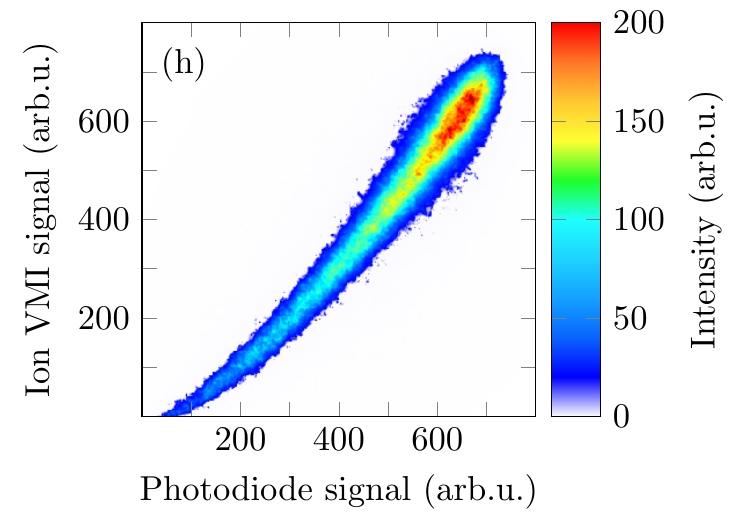}
&\includegraphics[width=0.3\linewidth]{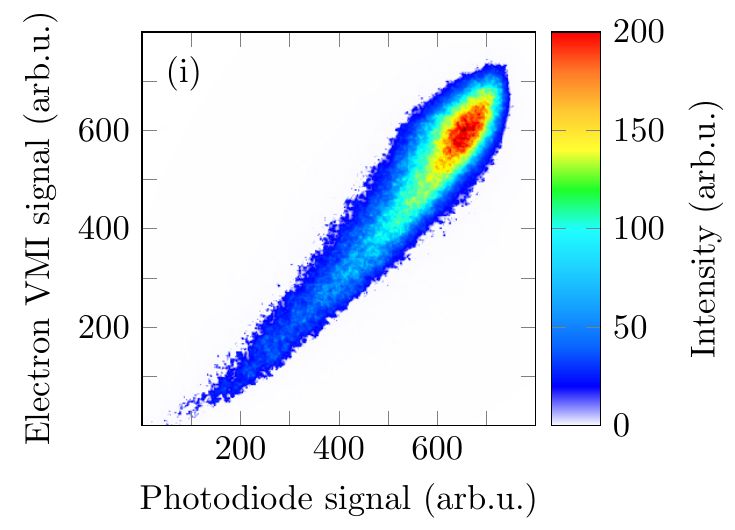}\\
\end{tabular} % We appreciate this elegant solution, but want to point out some aesthetically unpleasing details: panel f seems to stick out higher than panels d and e. Panel a seems to be wider than panel d.
%
%\hspace*{3.2mm}
%\subfigure{\includegraphics[width=0.30\textwidth]{Fig6/TOF}}
%\hspace*{-1.5mm}
%\subfigure{\includegraphics[width=0.31\textwidth]{Fig6/ionVMI}}
%\hspace*{2mm}
%\subfigure{\includegraphics[width=0.34\textwidth]{Fig6/electronVMI}}
%
%\subfigure{\includegraphics[width=0.30\textwidth]{Fig6/mass}}
%\hspace*{1mm}
%\subfigure{\includegraphics[width=0.30\textwidth]{Fig6/KIS}}
%\hspace*{2mm}
%\subfigure{\includegraphics[width=0.30\textwidth]{Fig6/KES_revised}}
%
%\subfigure{\includegraphics[width=0.32\textwidth]{Fig6/TOF-PD}}
%\subfigure{\includegraphics[width=0.32\textwidth]{Fig6/iVMI-PD}}
%\subfigure{\includegraphics[width=0.32\textwidth]{Fig6/eVMI-PD}}
\caption{Experimental results in the high resolution electron mode (2(b)) using high-order harmonics generated in argon to ionize N$_2$. (\textbf{a},\textbf{d},\textbf{g}) show ion TOF data ({a}), with the calibrated mass spectrum in ({d}) and the correlation between the total ion TOF signal and the XUV intensity measured by the XUV photodiode visualized by a color-coded scatter plot in ({g}). (\textbf{b},\textbf{e},\textbf{h}) show ion VMI data, with the average detector image (left part) and inverted image (right part) in~({b}). The~inverted images represent the reconstructed initial 3D momentum distributions obtained by inversion of the 2D projections recorded on the detector~\cite{VrakkingRSI2001}. The photoion kinetic energy spectrum is displayed in ({e}) with its attributed contributions shaded in different colors, calculated from the inverted image ({b}). The~correlation between the total ion VMI signal and the XUV intensity is exhibited in ({h}). ({c},{f},{i}) show electron VMI data, with the average detector image (left part) and inverted image (right part) in ({c}). The photoelectron kinetic energy spectrum is displayed in panel ({f}) with the attributed electronic molecular channels and the color-coded corresponding harmonic orders. The correlation between the total electron VMI signal and the XUV intensity is shown in (\textbf{i}). 
The slight bend observed for low intensities in the correlation plots is likely caused by a non-linear behavior of the photodiode in this signal region.}
   \label{fig:hhgmode2b}
\end{figure}

To be able to apply covariance analysis, synchronized single-shot data need to be acquired. At the high-intensity XUV beamline, the repetition rate is 10~Hz, and acquisition of single-shot data is relatively straightforward. However, the~intensity of the XUV light generated by laser-driven HHG typically varies on a shot-to-shot basis, introducing false correlations between channels that are physically not correlated. To circumvent this problem, a partial covariance method was used~\cite{Frasinski_1996}, through which the covariance maps were corrected for the fluctuating source intensity, recorded on a single-shot basis by an in-vacuum XUV photodiode placed after the focus of the XUV pulses.
 In this section, we describe the acquisition of data and the subsequent proof-of-principle application of ion-ion and ion-electron covariance mapping.

\subsection{Data Acquisition}

The successful implementation of the partial covariance analysis technique, discussed above, requires that the different single-shot observables can be reliably recorded in a synchronized fashion. In order to achieve this, a timestamp server was set up. After receiving the trigger from the laser master clock, the acquisition programs signal their readiness to the server. If all programs signal within a set timeout, the~server responds with a timestamp. Otherwise, it orders the programs to cancel the acquisition and to wait for the next trigger pulse. {Typically, about 5\% of the shots are discarded by the server.} To fully evaluate the performance of the instrument, the~acquisition of four synchronized signals is required:

\begin{itemize}
\item Ion time-of-flight trace
\item Ion velocity map image
\item Electron velocity map image
\item XUV intensity from photodiode
\end{itemize}

The ion TOF trace was obtained by decoupling the current from the back of the ion MCP and converting it into a voltage signal. The XUV photodiode generates a current, which is converted to a~voltage signal by means of a transimpedance amplifier. These analog ion TOF and XUV photodiode voltage traces are subsequently converted into digital signals using a two-channel, 1-GHz sampling rate, analog-to-digital converter (Agilent Acquiris DP1400), which~acquires 20,000~samples per laser shot in each of the channels, with a dynamic range of eight~bits. The ion and electron VMI data are recorded with two one-megapixel Allied Vision Technology Pike F-145B cameras that have an 8~bit dynamic range and a maximum frame rate of 30~Hz. In terms of storage rate, this amounts to more than 2~MB of data per shot, or~20~MB/s.

As a benchmark molecule, we used N$_2$. Figure~\ref{fig:hhgmode2b} shows simultaneously-acquired data from N$_2$, acquired using voltage ratios corresponding to Operation Mode 2(b), i.e., optimized for electron and ion VMI, but~with absolute voltages adapted for a maximum electron kinetic energy of 30~eV. The~data were recorded for 25,000 laser shots, corresponding to just above 40 minutes of acquisition time and amounting to $\sim$50~GB of data. 
The molecules were ionized using the harmonics 13--29 generated in argon, corresponding to photon energies between 20 and 45~eV, as displayed in Figure~\ref{fig:XUVspec}. 
The typical on-target pulse energy was $\sim$5~nJ, which~corresponds to approximately 10$^9$ photons per pulse over the whole bandwidth and 10$^8$ photons per pulse and harmonic. The count rates were on the order of 500 particles detected per shot for the ions, as well as the electrons.

Considering that the maximum available photon energy is just above the double ionization threshold of N$_2$ (42.88~eV~\cite{Hochlaf_1996}), a number of electronic states in N$_2$ with ionization potential (IP) in the range can be accessed: X$^2\Sigma _g^+$, A$^2\Pi _u$, B$^2\Sigma _u^+$, C$^2\Sigma _u^+$, F$^2\Sigma _g^+$ and H$^2\Sigma_g^+$. 
The average ion mass spectrum in Figure~\ref{fig:hhgmode2b}d exhibits two main features, which~were assigned to the N$^+$ and N$_2^+$ channels. While~the former also contains a contribution coming from stable N$_2^{2+}$ ions, only photons from the 29th harmonic or higher can contribute, and~considering the low intensity at these photon energies (Figure~\ref{fig:XUVspec}), the~contribution from this channel is expected to be negligible.~The mass resolution was estimated to $m/\Delta m=60$ for a mass of 28~u.

Figure~\ref{fig:hhgmode2b}b shows the photoion momentum distribution of all the ionic fragments that were generated in the interaction region, averaged over 25,000 shots. The anisotropy of the distribution, peaked in the vertical direction, is due to the fact that the cross-section for excitation to the dissociative states is higher for molecules aligned along the laser polarization direction. The faint horizontal ripple pattern in the images is a result of ions scattering on the mesh in the repeller electrode.
In Figure~\ref{fig:hhgmode2b}e, the corresponding photoion kinetic energy is displayed. It reveals three main features: a peaked contribution shaded in blue, a broader feature shaded in green and another broad contribution almost spreading out to the edge of the detector shaded in orange.
We attribute the strongest feature (shaded blue) to low kinetic energy N$_2^+$ ions and predissociation of the N$_2^+$ C$^2\Sigma _u^+$ state~\cite{Lucchini_2012}. The contribution shaded in green has previously been assigned to dissociation from the F$^2\Sigma _g^+$ state, and the features in the area shaded in orange are associated with dissociation from the highly-excited Rydberg-like states of N$_2^+$ and the H band~\cite{Eckstein_2015,Lucchini_2012,Aoto_2006}.

Figure \ref{fig:hhgmode2b}c,f shows the photoelectron momentum distribution and the photoelectron kinetic energy spectrum, respectively. From the photoelectron spectrum, it is apparent that the resolution of the spectrometer is limited by the bandwidth of the harmonics ($\sim$0.5~eV) rather than by the electron imaging. Figure~\ref{fig:hhgmode2b}f contains the assigned contributions originating from the X$^2\Sigma _g^+$, A$^2\Pi _u$ and B$^2\Sigma _u^+$ states, after taking their respective branching ratios, cross-sections \cite{Plummer_1977,Hammet_1976} and the relative XUV intensity at the photon energies contained in the XUV spectrum into account.~The ratio of the contributions is displayed by a stacked bar plot, where each state has a specific color assigned. The~much weaker contributions of the other cationic and dicationic states are not shown for clarity~\cite{Aoto_2006,Plummer_1977,Baltzer_1992,Ahmad_2006}. The~signal at kinetic energies smaller than 1~eV can be assigned to highly excited inner valence states of N$_2^+$ and possibly the first electronic states of N$_2^{2+}$~\cite{Eckstein_2015,Aoto_2006,Ahmad_2006}. 

As mentioned earlier, in order to extract information about the correlation between the different single-shot datasets, it is important that the sets are acquired simultaneously and can be analyzed in a synchronized fashion. That this is indeed the case for the acquired dataset readily verified by studying the color-coded scatter plots in Figure~\ref{fig:hhgmode2b}g--i, indicating a high degree of correlation between the XUV intensity measured by the photodiode and the other three detector signals.
\begin{figure}[H]
\centering
\includegraphics[width=0.5\linewidth]{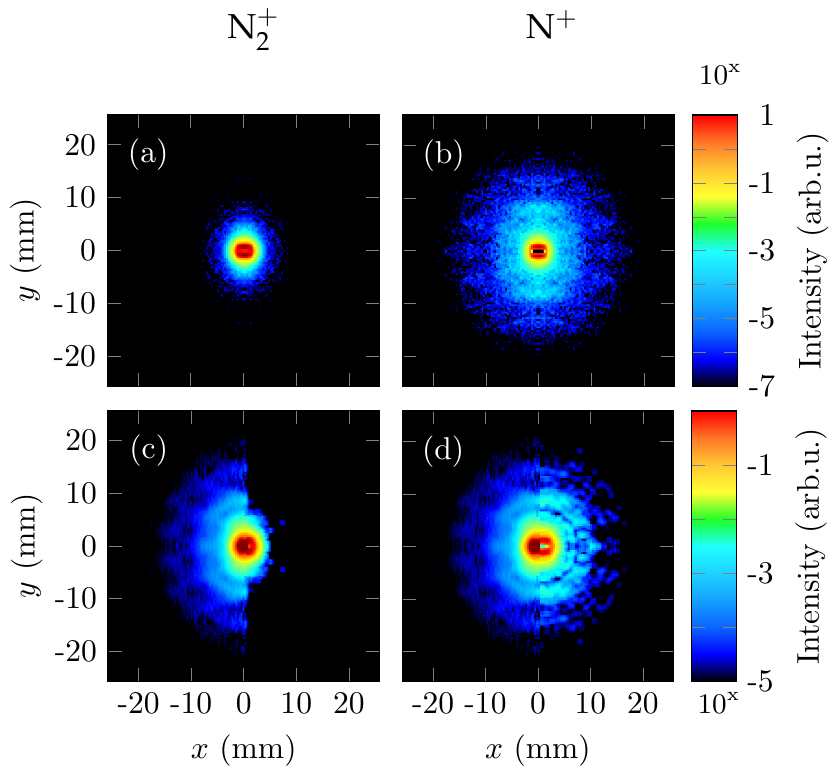}
%
%\hspace*{2.85cm}\normalsize N$_2^+$\hspace*{2.5cm}N$^+$
%\newline
%  	\hspace*{2.4cm}\begin{subfigure}{}   \includegraphics[width=0.255\textwidth]{Fig7/covarianceN2}\end{subfigure}		\hspace*{-3.6mm} \begin{subfigure}{}   \includegraphics[width=0.305\textwidth]{Fig7/covarianceN}  \end{subfigure}
%		\vspace*{-2.9mm}
%   \hspace*{22.2mm} 
%		\begin{subfigure}{}   \includegraphics[width=0.254\textwidth]{Fig7/covarianceN2inv}\end{subfigure}		\hspace*{-3.6mm} \begin{subfigure}{}   \includegraphics[width=0.304\textwidth]{Fig7/covarianceNinv}\end{subfigure}
 \caption{Results of ion-ion covariance analysis. (\textbf{a},\textbf{b}) show covariance maps of the N$_2^+$ and the N$^+$ ion TOF fragments correlated with the ion VMI detector image, respectively. Furthermore, these maps were treated like raw VMI images and inverted. The results are displayed in the right halves of (\textbf{c},\textbf{d}). For comparison the inverted image of the accumulated ion detector image from Figure~\ref{fig:hhgmode2b}b is reproduced in the left halves.}
   \label{fig:hhgcovIside}
 \end{figure}

\subsection{Covariance Analysis}

Once the synchronized acquisition of data from all the detectors is established, covariance analysis can be applied to different variables to reveal more information about the underlying physical processes during and after the photoionization.~As a proof-of-principle demonstration, we use covariance mapping to achieve post-acquisition mass selection of different ionic species. The aim is to extract the momentum distributions for individual ionic species from the average momentum distribution Figure~\ref{fig:hhgmode2b}b by calculating the covariance image with respect to the corresponding peak in the ion TOF spectrum shown in Figure~\ref{fig:hhgmode2b}a.~For mass selection, temporal windows around the two features in the time-of-flight spectrum in Figure~\ref{fig:hhgmode2b}a were chosen.~For this, the~partial covariance~\cite{Frasinski_1996}, $\text{pcov}(\mathbf{X},\mathbf{Y})$, between every pixel in the ion velocity map image ($\mathbf{X}$) with the total signal in the studied channel in the time-of-flight spectrum ($\mathbf{Y}$) was calculated. 

The resulting covariance maps are displayed in Figure~\ref{fig:hhgcovIside}a for the N$_2^+$ channel and in Figure~\ref{fig:hhgcovIside}b for the N$^+$ channel. As expected, the~N$_2^+$ map (a) is dominated by the zero kinetic energy parent ions from ionization to the X$^2\Sigma _g^+$, A$^2\Pi _u$ and B$^2\Sigma _u^+$ states. The N$^+$ map (b) shows correlation both at low kinetic energies, previously attributed to the predissociation of the N$_2^+$ C$^2\Sigma _u^+$ state and,~at higher kinetic energies, previously attributed to dissociation from the F$^2\Sigma _g^+$ state, highly excited Rydberg-like states of N$_2^+$ and the H band. Further, in order to compare it to the accumulated ion VMI data, the~same inversion procedure was applied to the covariance maps. The results are shown side by side with the reproduced inverted data from Figure~\ref{fig:hhgmode2b}b in the left halves of Figure~\ref{fig:hhgcovIside}c for N$_2^+$ and Figure~\ref{fig:hhgcovIside}d %please confirm if change it to “Figure~\ref{fig:hhgmode2b}c” and “Figure~\ref{fig:hhgmode2b}d” - good idea, but it refers to Figure 7 c and d
for N$^+$, with the right halves being the inverted covariance maps. These results clearly show that mass selection via covariance is a viable method to extract correlated information about the products of photoionization using the proposed instrument. 

\begin{figure}[H]\centering		 \includegraphics[]{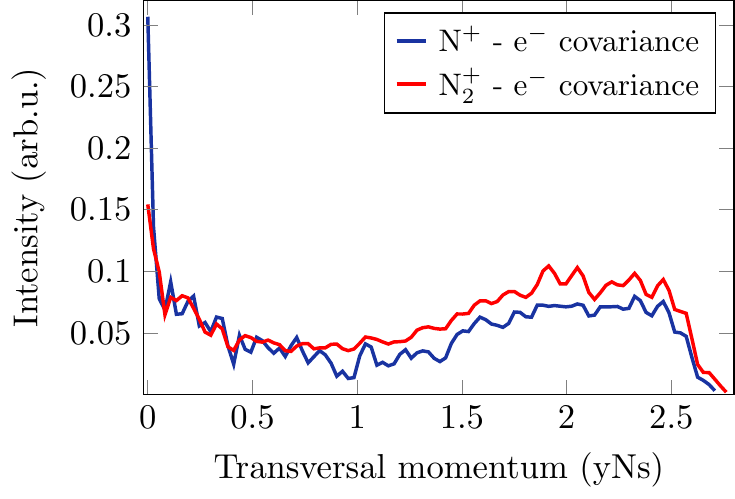}	
\caption{Results of ion-electron covariance analysis. Covariance between the photoelectron transversal momentum spectrum and the ion-TOF channels of N$^+$ (blue) and N$_2^+$ (red).~Qualitatively, the~N$^+$ channel shows stronger correlation with low momentum photoelectrons, whereas the N$_2^+$ channel exhibits a stronger correlation with photoelectrons at higher transversal momenta.}
\label{fig:elcov}\end{figure}

A natural next step is to consider ion-electron covariance using this scheme.~Consequently, the~photoelectron covariance maps were calculated in an analogous way to the ion covariance maps shown in Figure~\ref{fig:hhgcovIside}a,b.~Due to insufficient statistics, it was not possible to adequately invert the resulting covariance maps, and~instead, the results of an angular integration of these are shown in Figure~\ref{fig:elcov}. Since, as shown in Figure~\ref{fig:hhgmode2b}f, the~photoelectron spectrum primarily contains electrons from the X$^2\Sigma _g^+$, A$^2\Pi _u$ and B$^2\Sigma _u^+$ states that are all coupled to the N$_2^+$ channel, these completely dominate the covariance at photoelectron momenta between 1 and 2.6~yNs%define if appropriate - yNs is a SI unit (yocto=(10^-24) Newton second)
. While the individual peaks cannot be clearly resolved, partly due to a downsampling of the data for reasons of statistics, the~contribution is stronger for the N$_2^+$ channel in this regime, as expected. However, at close to zero momentum, the contribution associated with the production of N$^+$ from dissociation is dominating, which~is consistent with our previous attribution of these low kinetic energy electrons to excitation of inner valence states of N$_2^+$. Although qualitative, these first ion-electron covariance results indicate the feasibility of using covariance analysis to extract channel-specific photoelectron data from experiments with high single-shot count rates using the double VMIS presented here.

\section{Conclusions and Outlook}
\label{sec:conclusion}
In conclusion, we have reported on the design and performance of a velocity map imaging (VMI) spectrometer optimized for experiments using high-intensity extreme ultraviolet (XUV) sources such as laser-driven high-order harmonic generation (HHG) sources and free-electron lasers (FELs). The~instrument is versatile and allows for combining photo-electron and -ion detection modes, such~as ion time-of-flight (TOF), ion VMI and electron VMI, in different ways, depending on the required information and the process under study. The performance for the different detection modes was estimated using simulations, and~first experimental results from an intense HHG source were presented together with proof-of-principle application of covariance mapping to mass selection of photoions and extraction of channel-specific photoelectron spectra.

The analysis of the experimental results confirms that the acquired data contain information about the correlation between the products of the photoionization, and~the first covariance mapping analysis demonstrates the feasibility of extracting such information, indicating that the suggested approach is viable for extracting feature-rich correlated information from future experiments using high-intensity XUV sources.

\vspace{6pt}

\authorcontributions{Conceptualization: L.R. and P.J. Data curation: L.R., J.L. and S.M. Formal analysis: L.R., J.L., S.M. and P.J. Funding acquisition: P.J. Investigation: L.R., J.L., S.M., F.C., H.C.-A., B.O., J.P., H.W., P.R.~and~M.G. Project administration: P.J. Supervision: P.J.; Writing, original draft: L.R. and J.L. Writing, review and editing: L.R., J.L., S.M., F.C., H.C.-A., B.O., J.P., H.W., P.R., M.G. and P.J.}

%%%%%%%%%%%%%%%%%%%%%%%%%%%%%%%%%%%%%%%%%%
\funding{This research was supported by the Swedish Research Council, the~Swedish Foundation for Strategic Research and the Crafoord Foundation.~This project has received funding from the European Union's Horizon 2020 research and innovation program under the Marie Sklodowska-Curie Grant Agreement No.~641789~MEDEA.}%Please add: ``This research received no external funding'' or ``This research was funded by [name of funder] grant number [xxx].'' Check carefully that the details given are accurate and use the standard spelling of funding agency names at \url{https://search.crossref.org/funding}, any errors may affect your future funding. - we had that in the acknowledgement previously.

%\acknowledgments{}

\conflictofinterests{The authors declare no conflict of interest. The founding sponsors had no role in the design of the study; in the collection, analyses or~interpretation of data; in the writing of the manuscript; nor~in the decision to publish the results.} 

%\bibliographystyle{mdpi}
%\renewcommand\bibname{References}
%bibliography{Ref_lib,additional}
\reftitle{References}

\end{document}